\documentclass[aps,prd,twocolumn,superscriptaddress]{revtex4}
\usepackage{epsfig,epsf}
\usepackage{amsmath}
\usepackage{amsthm}
\usepackage{amsfonts}
\usepackage{amssymb}
\usepackage{dsfont}
\usepackage{multirow}
\usepackage{appendix}
\usepackage{slashed}
\usepackage[active]{srcltx}
\usepackage{psfrag}

\begin{document}

\title{Fully-beauty tensor tetraquark}
\date{\today }
\author{S.~S.~Agaev}
\affiliation{Institute for Physical Problems, Baku State University, Az--1148 Baku,
Azerbaijan}
\author{K.~Azizi}
\thanks{Corresponding author: kazem.azizi@ut.ac.ir}
\affiliation{Department of Physics, University of Tehran, North Karegar Avenue, Tehran
14395-547, Iran}
\affiliation{Department of Physics, Faculty of Engineering and Natural Sciences, Dogus
University, Dudullu-\"{U}mraniye, 34775 Istanbul, T\"{u}rkiye}
\author{H.~Sundu}
\affiliation{Department of Physics Engineering, Istanbul Medeniyet University, 34700
Istanbul, T\"{u}rkiye}

\begin{abstract}
Parameters of the fully-beauty tensor tetraquark $T=bb\overline{b}\overline{b%
}$ are computed using QCD sum rule method. The mass $m$ and coupling $%
\Lambda $ of this state are evaluated by means of two-point sum rule
approach. Prediction $m=(18530 \pm 86)~\mathrm{MeV}$ for the mass
demonstrates that $T$ is stable against decays to pairs of $\eta_b \eta_b$
and $\Upsilon \Upsilon$ mesons. But it transforms to conventional particles
due to $\overline{b}b$ annihilations to light quarks $\overline{q} q $ and $%
\overline{s}s$ followed by creation of mesons $B^{(\ast)+}B^{( \ast)-}$, $%
B^{(\ast)0}\overline{B}^{(\ast)0}$ and $B_s^{(\ast)0}\overline{B}
_s^{(\ast)0}$. Partial width of these decay channels are calculated by
invoking technical tools of three-point sum rule method. The latter allows
us to estimate strong couplings at relevant tetraquark-meson-meson vertices
and, hence, a width of the process under consideration. Our prediction $%
\Gamma=(48 \pm 6)~\mathrm{MeV}$ along with the mass of this state can be
valuable for experimental studies of all-heavy resonances.
\end{abstract}

\maketitle


\section{Introduction}

\label{sec:Int} 

Investigations of multiquark hadrons which cannot be classified as
conventional mesons or baryons have a long history. Thus assumption about
four-quark nature of light mesons from the lowest scalar nonet was made
already in Ref.\ \cite{Jaffe:1976ig}. Another interesting result is
connected with a state built of six light quarks which if exists would be
strong-interaction stable particle \cite{Jaffe:1976yi}.

Exotic mesons composed of heavy and light quarks were also objects for
theoretical studies. Structures containing heavy $QQ$ diquarks ($Q=c$ or $b$
) and light antidiquarks $\overline{q}\overline{q}^{\prime }$ were
considered as real candidates to multiquark hadrons stable against strong
and/or electromagnetic decays. In pioneering publications \cite%
{Ader:1981db,Lipkin:1986dw,Zouzou:1986qh} it was shown that structures $QQ%
\overline{q}\overline{q}^{\prime }$ may be stable particles if the ratio $%
m_{Q}/m_{q}$ is large. Recent explicit calculations confirmed this
assumption and led to conclusions that the axial-vector tetraquark $bb%
\overline{u}\overline{d}$ is probably such a state \cite{Karliner:2017qjm}.
Strong-interaction stable nature of the fully-heavy tetraquarks with
different contents and spin-parities were revealed in Ref.\ \cite%
{Eichten:2017ffp} as well.

Naturally, these tetraquarks can decay through weak processes. In our
articles \cite%
{Agaev:2018khe,Agaev:2020mqq,Agaev:2020dba,Agaev:2019kkz,Sundu:2019feu,
Agaev:2019lwh,Agaev:2020zag} we evaluated full widths of the scalar and
axial-vector tetraquarks $bb\overline{u}\overline{d}$, $bb\overline{u}%
\overline{s}$ and $bc\overline{u}\overline{d}$ \ by exploring their various
semileptonic and nonleptonic decay channels. These exotic mesons may be
observed soon in ongoing and future experiments.

Experimental situation with particles $QQ\overline{Q}\overline{Q}$ is more
promising. In fact, the LHCb-ATLAS-CMS collaborations discovered four $X$
resonances with masses $6.2-7.2~\mathrm{GeV}$ in the di-$J/\psi $ and $%
J/\psi \psi ^{\prime }$ mass distributions. Measured parameters of these
resonances provide valuable information on fully-heavy exotic mesons \cite%
{LHCb:2020bwg,ATLAS:2023bft,CMS:2023owd,CMS:2025fpt,CMS:2026tiu}. This is
connected with the fact that $X$ structures are supposedly tetraquarks $cc%
\overline{c}\overline{c}$ and their studies can shed light on features of
fully-heavy systems. There are observed processes with $b$-mesons at the
final states as well. In fact, decays to $J/\psi \Upsilon $ and $\Upsilon
\Upsilon $ mesons were seen and studied by D0 and CMS experiments \cite%
{D0:2015dyx,CMS:2016liw}. Such final products imply generation of
intermediate $cc\overline{c}\overline{c}$, $bc\overline{b}\overline{c}$ and $%
bb\overline{b}\overline{b}$ structures followed by transformations to
conventional heavy mesons. In other words, there are data confirming
existence of fully-beauty tetraquarks $bb\overline{b}\overline{b}$, which
deserve detailed analysis.

Theoretical examinations of fully-heavy exotic mesons were actually
performed in numerous publications \cite%
{Berezhnoy:2011xn,Karliner:2016zzc,Wu:2016vtq,Chen:2016jxd,Wang:2017jtz,Richard:2017vry,Esposito:2018cwh}
in which their parameters were investigated by means of different methods
and schemes. Discoveries of LHCb-ATLAS-CMS collaborations triggered
appearance of new interesting articles where authors addressed various
problems of \ fully-heavy tetraquarks \cite%
{Wang:2022xja,Faustov:2022mvs,Niu:2022vqp,Dong:2022sef,Yu:2022lak,An:2022qpt,Kuang:2023vac,Liu:2020eha,Malekhosseini:2025hyx,Likene:2026pga}%
. The $X$ resonances were investigated in Refs.\ \cite%
{Agaev:2023wua,Agaev:2023ruu,Agaev:2023gaq,Agaev:2023rpj} using QCD sum rule
(SR) method \cite{Shifman:1978bx,Shifman:1978by}. In these works, we
considered the $X$ structures as scalar particles using both
diquark-antidiquark and hadronic molecule models and calculated masses and
decay widths of these states. But new measurements of CMS experiment
demonstrated that these structures are tensor states $J^{\mathrm{PC}}=2^{++}$%
. As tensor particles they were studied in our articles \cite%
{Agaev:2026mif,Agaev:2026izr} as well.

Structures $bb\overline{b}\overline{b}$ were objects of interesting
investigations in which authors applied different methods. For instance, the
masses of fully- heavy tetraquarks may be found by solving nonrelativistic
Schrodinger equation \cite{Berezhnoy:2011xn}. In accordance with this
article all fully-beauty (i.e., ones with $J=0$, $1,2$) tetraquarks reside
below $\Upsilon \Upsilon $ threshold, and cannot be seen in this mass
distribution. The scalar tetraquark $bb\overline{b}\overline{b}$ was
explored also in Ref.\ \cite{Karliner:2016zzc}. Prediction for $m_{\mathrm{4b%
}}=(18826\pm 25)~\mathrm{MeV}$ permitted the authors to conclude that this
particle does not decay to a pair of mesons $\Upsilon \Upsilon $, whereas
decay to a $\eta _{b}\eta _{b}$ pair is its allowed decay mode. In Ref. \cite%
{Chen:2016jxd} the masses of tetraquarks $bb\overline{b}\overline{b}$ with
different spin-parities were computed using SR method. The analysis revealed
that that the masses of the scalar structures vary within limits $%
18.45-18.59\ \mathrm{GeV}$. As a result, the scalar particle is stable
against strong dissociation to the hidden-beauty mesons.

The scalar $bb\overline{b}\overline{b}$ tetraquarks $X_{\mathrm{4b}}$ and $%
T_{\mathrm{4b}}$ were also considered in our articles \cite%
{Agaev:2023wua,Agaev:2023gaq,Agaev:2023ara}. The tetraquark $X_{\mathrm{4b}}$
was built of axial-vector diquarks, whereas to construct $T_{\mathrm{4b}}$
we used pseudoscalar components. We estimated their masses in the range $%
(18540\pm 50)~\mathrm{MeV}$ and $(18858\pm 50)~\mathrm{MeV}$, respectively.
These results indicate that $X_{\mathrm{4b}}$ does not decay through
channels $X_{\mathrm{4b}}\rightarrow \eta _{b}\eta _{b}$ and $X_{\mathrm{4b}%
}\rightarrow \Upsilon \Upsilon $. At the same time, $T_{\mathrm{4b}}$ falls
apart to a pair of $\eta _{b}\eta _{b}$ mesons. But this does not mean that $%
X_{\mathrm{4b}}$ is stable against all strong decays. In fact, annihilations
of $b\overline{b}$ quarks to light quark pairs $q\overline{q}$ and $s%
\overline{s}$ open new channels for transformation of $X_{\mathrm{4b}}$ to
conventional mesons through strong processes \cite%
{Becchi:2020mjz,Becchi:2020uvq}. This alternative mechanism for strong
decays was employed to estimate the width of $X_{\mathrm{4b}}$ in Ref.\ \cite%
{Agaev:2023ara}.

In present paper, we are going to investigate the fully-beauty tensor
structure $\mathrm{T}=bb\overline{b}\overline{b}$ in the diquark-antidiquark
model. We compute the mass and decay width of this state. Our analysis is
performed in the framework of SR approach and shows that the tetraquark $%
\mathrm{T}$ is stable against fall-apart channels. We evaluate its decay
width by exploring the processes $\mathrm{T}\rightarrow B^{(\ast )+}B^{(\ast
)-}$, $B^{(\ast )0}\overline{B}^{(\ast )0}$ and $\mathrm{T}\rightarrow
B_{s}^{(\ast )0}\overline{B}_{s}^{(\ast )0}$ generated by aforementioned
mechanism. To estimate the partial widths of these modes, we employ the
three-point SR method, which is necessary to find strong form factors $%
g_{i}(q^{2})$ describing strong interaction at relevant
tetraquark-meson-meson vertices. The strong couplings $g_{i}$ determined
after extrapolation of $g_{i}(q^{2})$ to the mass-shell of a final meson are
utilized to compute widths of the corresponding decay modes and infer the
full width $\Gamma (\mathrm{T})$ of the tetraquark.

This article is organized in the following manner: In Section \ref{sec:Mass}%
, we determine the mass and current coupling of the tetraquark $\mathrm{T}
$. Strong decays $\mathrm{T}\rightarrow B^{(\ast )+}B^{(\ast )-}$, $B^{(\ast
)0}\overline{B}^{(\ast )0}$ are studied in Sec.\ \ref{sec:Decays1}. Partial
widths of the modes $\mathrm{T}\rightarrow B_{s}^{(\ast )0}\overline{B}%
_{s}^{(\ast )0}$ are calculated in Sec.\ \ref{sec:Decays2}. In this section,
we estimate also the full width of the tetraquark $\mathrm{T}$. Last section
is reserved for our concluding notes.


\section{Mass and current coupling of the tetraquark $\mathrm{T}$}

\label{sec:Mass}%

The mass $m$ and current coupling $\Lambda $ of the tensor tetraquark $%
\mathrm{T}$ can be evaluated using expressions obtained in the context of
QCD sum rule method \cite{Shifman:1978bx,Shifman:1978by}.

For this purpose we consider the two-point correlation function
\begin{equation}
\Pi _{\mu \nu \alpha \beta }(p)=i\int d^{4}xe^{ipx}\langle 0|\mathcal{T}%
\{I_{\mu \nu }(x)I_{\alpha \beta }^{\dag }(0)\}|0\rangle ,  \label{eq:CF1}
\end{equation}%
where $I_{\mu \nu }(x)$ is the interpolating current for the tensor state $%
\mathrm{T}$.

We model $\mathrm{T}$ as a structure built of axial-vector diquark and
antidiquark components, therefore $I_{\mu \nu }(x)$ has the following form
\begin{eqnarray}
I_{\mu \nu }(x) &=&b_{a}^{T}(x)C\gamma _{\mu }b_{b}(x)\overline{b}%
_{a}(x)\gamma _{\nu }C\overline{b}_{b}^{T}(x)  \notag \\
&&+b_{a}^{T}(x)C\gamma _{\nu }b_{b}(x)\overline{b}_{a}(x)\gamma _{\mu }C%
\overline{b}_{b}^{T}(x),  \label{eq:CR}
\end{eqnarray}%
where $b(x)$ is $b$-quark field with $a$ and $b$ being the color indices,
and $C$ is the charge conjugation matrix. The current $I_{\mu \nu }(x)$
describes the particle with the spin-parities $J^{\mathrm{PC}}=2^{++}$.

The current in Eq.\ (\ref{eq:CR}) is explicitly symmetric in its Lorentz
indices, but its tracelessness is not imposed at the level of the current
itself. It can therefore, in principle, contain a trace component which may
couple to states with quantum numbers different from those of the desired
spin-2 state. However, this trace component does not contribute to the
particular Lorentz structure used to derive the mass and current coupling
sum rules in the present analysis (see, Eq.\ (\ref{eq:Phys2}) below).  As a result, an explicit
 trace term in the interpolating current is not necessary for
the sum rules considered here.

To find SRs for the parameters $m$ and $\Lambda $, one should compute the
correlation function $\Pi _{\mu \nu \alpha \beta }(p)$ using two approaches.
In the first one, the correlator is expressed in terms of physical
parameters of the tetraquark which leads to physical component $\Pi _{\mu
\nu \alpha \beta }^{\mathrm{Phys}}(p)$ of SRs. To this end, we insert into $%
\Pi _{\mu \nu \alpha \beta }(p)$ a complete set of intermediate states, and
carry out the integration over $x$. As a result, we get
\begin{eqnarray}
\Pi _{\mu \nu \alpha \beta }^{\mathrm{Phys}}(p) &=&\frac{\langle 0|I_{\mu
\nu }|\mathrm{T}(p,\epsilon )\rangle \langle \mathrm{T}(p,\epsilon
)|I_{\alpha \beta }^{\dag }|0\rangle }{m^{2}-p^{2}}  \notag \\
&&+\cdots ,
\end{eqnarray}%
where $\epsilon =\epsilon _{\mu \nu }(p)$ is the polarization tensor of the
tetraquark $\mathrm{T}$. Above, the term that corresponds to ground-level
particle $\mathrm{T}$ is presented explicitly, whereas contributions of
higher resonances and continuum states are shown by the ellipses.

The expression for $\Pi _{\mu \nu \alpha \beta }^{\mathrm{Phys}}(p)$ can be
simplified by employing the matrix element
\begin{equation}
\langle 0|I_{\mu \nu }|\mathrm{T}(p,\epsilon (p))\rangle =\Lambda \epsilon
_{\mu \nu }(p).  \label{eq:ME1}
\end{equation}%
Having used Eq.\ (\ref{eq:ME1}) in the correlator $\Pi _{\mu \nu \alpha
\beta }^{\mathrm{Phys}}(p)$ and performed required operations, one finds
\begin{eqnarray}
\Pi _{\mu \nu \alpha \beta }^{\mathrm{Phys}}(p) &=&\frac{\Lambda ^{2}}{%
m^{2}-p^{2}}\left\{ \frac{1}{2}\left( g_{\mu \alpha }g_{\nu \beta }+g_{\mu
\beta }g_{\nu \alpha }\right) \right.  \notag \\
&&\left. +\text{ other components}\right\} +\cdots .  \label{eq:Phys2}
\end{eqnarray}%
The function $\Pi _{\mu \nu \alpha \beta }^{\mathrm{Phys}}(p)$ is composed
of contributions with different Lorentz structures. The term $\sim (g_{\mu
\alpha }g_{\nu \beta }+g_{\mu \beta }g_{\nu \alpha })$ arises from a spin-$2$
particle. Therefore, it is convenient to employ in our analysis this
contribution and related invariant amplitude $\Pi ^{\mathrm{Phys}}(p^{2})$.

To find the QCD side $\Pi _{\mu \nu \alpha \beta }^{\mathrm{OPE}}(p)$ of SRs
we use in $\Pi _{\mu \nu \alpha \beta }(p)$ the current $I_{\mu \nu }(x)$
and contract quark fields. We get
\begin{eqnarray}
&&\Pi _{\mu \nu \alpha \beta }^{\mathrm{OPE}}(p)=i\int d^{4}xe^{ipx}\left\{
\left\{ \mathrm{Tr}\left[ \gamma _{\nu }\widetilde{S}_{b}^{b^{\prime
}b}(-x)\gamma _{\beta }S_{b}^{a^{\prime }a}(-x)\right] \right. \right.
\notag \\
&&\left. -\mathrm{Tr}\left[ \gamma _{\nu }\widetilde{S}_{b}^{a^{\prime
}b}(-x)\gamma _{\beta }S_{b}^{b^{\prime }a}(-x)\right] \right\} \left\{
\mathrm{Tr}\left[ \gamma _{\alpha }\widetilde{S}_{b}^{a^{\prime }a}(x)\gamma
_{\mu }\right. \right.  \notag \\
&&\left. \left. \times S_{b}^{bb^{\prime }}(x)\right] -\mathrm{Tr}\left[
\gamma _{\alpha }\widetilde{S}_{b}^{ba^{\prime }}(x)\gamma _{\mu
}S_{b}^{ab^{\prime }}(x)\right] \right\} +\left( \mu \leftrightarrow \nu
\right)  \notag \\
&&\left. +\left( \alpha \leftrightarrow \beta \right) +\left( \mu
\leftrightarrow \nu ,\alpha \leftrightarrow \beta \right) \right\} ,
\label{eq:QCD1}
\end{eqnarray}%
where $S_{b}(x)$ is $b$-quark propagator 
\begin{eqnarray}
&&S_{b}^{ab}(x)=\frac{i}{(2\pi )^{4}}\int d^{4}ke^{-ikx}\Bigg \{\frac{\delta
_{ab}\left( {\slashed k}+m_{b}\right) }{k^{2}-m_{b}^{2}}  \notag \\
&&-\frac{g_{s}G_{ab}^{\alpha \beta }}{4}\frac{\sigma _{\alpha \beta }\left( {%
\slashed k}+m_{b}\right) +\left( {\slashed k}+m_{b}\right) \sigma _{\alpha
\beta }}{(k^{2}-m_{b}^{2})^{2}}  \notag \\
&&+\frac{g_{s}^{2}G^{2}}{12}\delta _{ab}m_{b}\frac{k^{2}+m_{b}{\slashed k}}{%
(k^{2}-m_{b}^{2})^{4}}+\cdots \Bigg \}.  \label{eq:HQprop}
\end{eqnarray}%
Above, we have used the notation
\begin{equation}
G_{ab}^{\alpha \beta }\equiv G_{A}^{\alpha \beta }\lambda _{ab}^{A}/2,
\end{equation}%
where $G_{A}^{\alpha \beta }$ is the gluon field-strength tensor, and $%
\lambda ^{A},\ $ $A=1,2,\ldots 8$ are the Gell-Mann matrices. The propagator
$\widetilde{S}_{b}(x)$ in Eq. (\ref{eq:QCD1}) is defined by the formula%
\begin{equation}
\widetilde{S}_{b}(x)=CS_{b}^{T}(x)C.
\end{equation}

The correlator $\Pi _{\mu \nu \alpha \beta }^{\mathrm{OPE}}(p)$ should be
calculated with some accuracy by employing operator product expansion ($%
\mathrm{OPE}$). After extracting in $\Pi _{\mu \nu \alpha \beta }^{\mathrm{%
OPE}}(p)$ a contribution proportional to $(g_{\mu \alpha }g_{\nu \beta
}+g_{\mu \beta }g_{\nu \alpha })$ and labeling by $\Pi ^{\mathrm{OPE}%
}(p^{2}) $ the corresponding amplitude, we determine desired SRs. To this
end, we equate amplitudes $\Pi ^{\mathrm{Phys}}(p^{2})$ and $\Pi ^{\mathrm{%
OPE}}(p^{2})$ and perform usual manipulations of SR approach. Stated
differently, we apply the Borel transformation to suppress effects of higher
resonances and continuum states. Later, in the framework of the quark-hadron
duality assumption, we subtract these contributions from QCD side of the SR
equality. These operations transform $\Pi ^{\mathrm{OPE}}(p^{2})$ to $\Pi
(M^{2},s_{0})$ which is a function of the Borel and continuum subtraction
parameters $M^{2}$ and $s_{0}$. The SRs for the mass $m$ and current
coupling $\Lambda $ are given by the formulas
\begin{equation}
m^{2}=\frac{\Pi ^{\prime }(M^{2},s_{0})}{\Pi (M^{2},s_{0})},  \label{eq:Mass}
\end{equation}%
and
\begin{equation}
\Lambda ^{2}=e^{m^{2}/M^{2}}\Pi (M^{2},s_{0}),  \label{eq:Coupl}
\end{equation}%
where $\Pi ^{\prime }(M^{2},s_{0})=d\Pi (M^{2},s_{0})/d(-1/M^{2})$.

The transformed amplitude $\Pi (M^{2},s_{0})$ has the form%
\begin{equation}
\Pi (M^{2},s_{0})=\int_{16m_{b}^{2}}^{s_{0}}ds\rho ^{\mathrm{OPE}%
}(s)e^{-s/M^{2}}+\Pi (M^{2}).  \label{eq:CorrF}
\end{equation}%
We calculate $\Pi (M^{2},s_{0})$ by including into analysis dimension-$4$
terms $\sim \langle \alpha _{s}G^{2}/\pi \rangle $. In Eq.\ (\ref{eq:CorrF})
$\rho ^{\mathrm{OPE}}(s)$ is the spectral density which is equal to the
imaginary part of the amplitude $\Pi ^{\mathrm{OPE}}(p^{2})$. The
contribution $\Pi (M^{2})$ is obtained directly from $\Pi ^{\mathrm{OPE}%
}(p^{2})$ and contains terms absent in $\rho ^{\mathrm{OPE}}(s)$. Note that
analytical expressions of the functions $\rho ^{\mathrm{OPE}}(s)$ and $\Pi
(M^{2})$ are cumbersome and not provided here.

In numerical analysis one has to fix parameters which enter to the sum
rules. As the mass $m_{b}$ of $b$ quark and gluon condensate $\langle \alpha
_{s}G^{2}/\pi \rangle $ we employ%
\begin{eqnarray}
&&m_{b}=(4.183\pm 0.007)~\mathrm{GeV},  \notag \\
&&\langle \alpha _{s}G^{2}/\pi \rangle =(0.012\pm 0.004)~\mathrm{GeV}^{4},
\end{eqnarray}%
which are universal entries.
\begin{figure}[tbp]
\includegraphics[width=8.5cm]{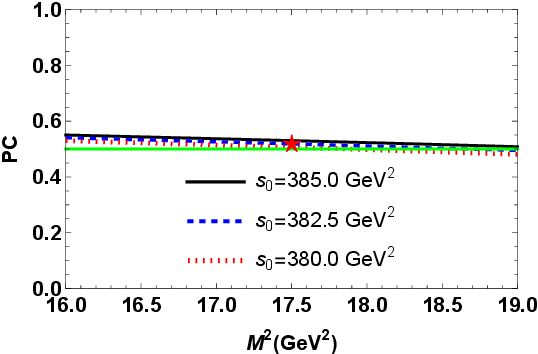}
\caption{$\mathrm{PC}$ as a function of $M^{2}$ at some $s_{0}$. The
horizontal line represents the limit $\mathrm{PC}=0.5$. The star fixes the
point $M^{2}=17.5~\mathrm{GeV}^{2},s_{0}=382.5~\mathrm{GeV}^{2}$. }
\label{fig:PC}
\end{figure}

The quantities $M^{2}$ and $s_{0}$ depend on a process under investigation
and must meet usual restrictions of SR analysis. They include dominance of
pole contribution ($\mathrm{PC}$) in physical quantities, convergence of $%
\mathrm{OPE}$ and slight dependence of $m$ and $\Lambda $ on $M^{2}$ and $%
s_{0}$: These constraints are important for soundness of SR predictions.
Therefore, we require satisfaction of $\mathrm{PC}\geq 0.5$, where
\begin{equation}
\mathrm{PC}=\frac{\Pi (M^{2},s_{0})}{\Pi (M^{2},\infty )}.  \label{eq:PC}
\end{equation}

Because $\Pi (M^{2},s_{0})$ contains the perturbative and dimension-$4$
contribution $\Pi ^{\mathrm{Dim4}}(M^{2},s_{0})$, to ensure convergence of $%
\mathrm{OPE}$, we impose the constraint $|\Pi ^{\mathrm{Dim4}%
}(M^{2},s_{0})|\leq 0.05|\Pi (M^{2},s_{0})|$. It is worth to emphasize that
these two constraints permit one to get maximal and minimal values of $M^{2}$%
, respectively.

Our computations demonstrate that regions for $M^{2}$ and $s_{0}$
\begin{equation}
M^{2}\in \lbrack 16,19]~\mathrm{GeV}^{2},\ s_{0}\in \lbrack 380,385]~\mathrm{%
GeV}^{2},  \label{eq:Wind1}
\end{equation}%
satisfy all required conditions. In fact, the averaged over $s_{0}$ pole
contribution is $\mathrm{PC}\approx 0.49$ and $\mathrm{PC}\approx 0.54$ at $%
19~\mathrm{GeV}^{2}$ and $16~\mathrm{GeV}^{2}$, respectively. The term $|\Pi
^{\mathrm{Dim4}}(M^{2},s_{0})|$ at $M^{2}=16~\mathrm{GeV}^{2}$ is around $%
1\% $ of the amplitude $\Pi (M^{2},s_{0})$. We show in Fig.\ \ref{fig:PC}
the $\mathrm{PC}$ as a function of the Borel parameter $M^{2}$.

\begin{widetext}

\begin{figure}[htbp]
\begin{center}
\includegraphics[totalheight=6cm,width=8cm]{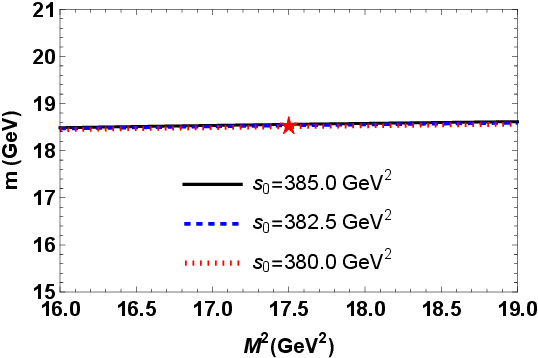}
\includegraphics[totalheight=6cm,width=8cm]{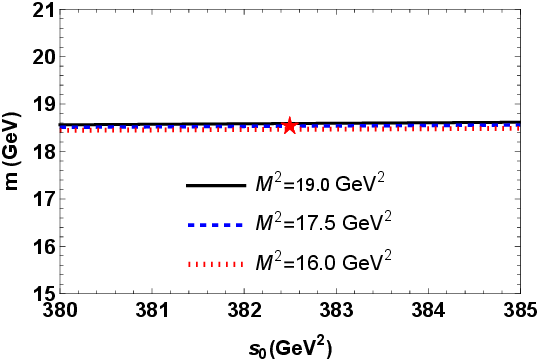}
\end{center}
\caption{The mass $m$ as a function of the parameters $M^{2}$ (left), and $s_0$ (right).}
\label{fig:Mass}
\end{figure}

\end{widetext}

We calculate $m$ and $\Lambda $ in the windows Eq.\ (\ref{eq:Wind1}) and
find their average values
\begin{eqnarray}
&&m=(18530\pm 86)~\mathrm{MeV},  \notag \\
&&\Lambda =(14.11\pm 1.44)~\mathrm{GeV}^{5}.  \label{eq:Result1}
\end{eqnarray}%
The values in Eq.\ (\ref{eq:Result1}) are corresponding to the SR results at
the point $M^{2}=17.5~\mathrm{GeV}^{2}$ and $s_{0}=382.5~\mathrm{GeV}^{2}$,
where $\mathrm{PC}\approx 0.51$, which guaranties dominance of $\mathrm{PC}$
in the quantities $m$ and $\Lambda $. Ambiguities in Eq.\ (\ref{eq:Result1})
are connected with those in $M^{2}$ and $s_{0}$: The quark mass $m_{b}$ and
condensate $\langle \alpha _{s}G^{2}/\pi \rangle $ practically do not create
sizable errors. Thus, errors in $m$ are equal to $\pm 0.5\%$ , while in the
case of $\Lambda $ they are $\pm 10.2\%$. These theoretical uncertainties
are within standard limits of SR analysis and prove correctness of the
extracted predictions. Variations of the mass $m$ on the parameters $M^{2}$
and $s_{0}$ are shown in Fig.\ \ref{fig:Mass}.

\section{Decays $\mathrm{T}\rightarrow B^{(\ast )+}B^{(\ast )-}$, $B^{(\ast
)0}\overline{B}^{(\ast )0-}$}

\label{sec:Decays1}

\begin{figure}[h]
\includegraphics[width=8.5cm]{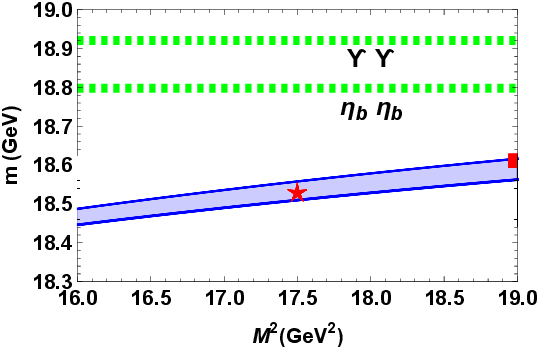}
\caption{The mass $m$ of the state $\mathrm{T}$ in used domains for the
parameters $M^{2}$ and $s_{0}$. The masses $m=18530~\mathrm{MeV}$ and $%
m=18616~\mathrm{MeV}$ are denoted by the star and rectangle, respectively.
The two-meson thresholds are plotted as dashed lines.}
\label{fig:Thresholds}
\end{figure}

Prediction for the mass $m$ determines possible decay mechanism of the tetraquark $\mathrm{T}$. Thus, there
are two important thresholds for fully-beauty tetraquarks, i.e., $2\eta _{b}$
and $2\Upsilon $ thresholds $18798~\mathrm{MeV}$ and $18921~\mathrm{MeV}$,
respectively. Possible decay modes of \ the tensor tetraquark $\mathrm{T}$
to ordinary mesons are limited by its position in this mass scale. Even in
the maximally allowed value $m=18616~\mathrm{MeV}$ it is below $2\eta _{b}$
and $2\Upsilon $ thresholds and stable against strong decays to conventional
$b\overline{b}$ mesons (see, Fig.\ \ref{fig:Thresholds}). Such structures
transform to known particles owing to annihilation of $b\overline{b}$ to
light quark-antiquark pairs which later convert to heavy-light mesons \cite%
{Becchi:2020mjz}.

The tetraquark $\mathrm{T}$ decays to ordinary mesons due to annihilation of
$b\overline{b}$ quarks to light quark-antiquark pairs \cite%
{Becchi:2020mjz,Becchi:2020uvq,Agaev:2023ara} and generation of $BB$ mesons
with appropriate quantum numbers. Here, we consider the processes $\mathrm{T}%
\rightarrow B^{+}B^{-}$,$\ B^{0}\overline{B}^{0}$, $B^{^{\ast }+}B^{\ast -}$%
, and $B^{\ast 0}\overline{B}^{\ast 0}$.

It is worth to emphasize that correlation functions of, for example,
processes $\mathrm{T}\rightarrow B^{+}B^{-}$ and $\mathrm{T}\rightarrow B^{0}%
\overline{B}^{0}\ $ differ from each other only by propagators of $u$ and $d$
quarks. Since we apply the simplification $m_{u}=m_{d}=0$, and also neglect
the tiny numerical differences in the masses of the charged and neutral $B$
mesons, processes $\mathrm{T}\rightarrow B^{+}B^{-}$ and $\mathrm{T}%
\rightarrow B^{0}\overline{B}^{0}\ $ have the same widths. This
approximation is correct also for the decays $\mathrm{T}\rightarrow $ $%
B^{^{\ast }+}B^{\ast -}$ and $B^{\ast 0}\overline{B}^{\ast 0}$.


\subsection{Processes $\mathrm{T}\rightarrow B^{+}B^{-}$,$\ B^{0}\overline{B}%
^{0}$}

The channel $\mathrm{T}\rightarrow B^{+}B^{-}$ can be analyzed using the
following correlation function
\begin{eqnarray}
\Pi _{\mu \nu }(p,p^{\prime }) &=&i^{2}\int d^{4}xd^{4}ye^{ip^{\prime
}y}e^{-ipx}\langle 0|\mathcal{T}\{J^{B^{+}}(y)  \notag \\
&&\times J^{B^{-}}(0)I_{\mu \nu }^{\dagger }(x)\}|0\rangle ,
\end{eqnarray}%
where currents $J^{B^{+}}(x)$ and $J^{B^{-}}(x)$ are given by the expressions%
\begin{equation}
J^{B^{+}}(x)=\overline{b}_{i}(x)i\gamma _{5}u_{i}(x),\text{ }J^{B^{-}}(x)=%
\overline{u}_{j}(x)i\gamma _{5}b_{j}(x).
\end{equation}%
The SR for the form factor $g_{1}(q^{2})$ that explains the strong
interaction of hadrons at the vertex $\mathrm{T}B^{+}B^{-}$ is acquired by
evaluating the correlators $\Pi _{\mu \nu }^{\mathrm{Phys}}(p,p^{\prime })$
and $\Pi _{\mu \nu }^{\mathrm{OPE}}(p,p^{\prime })$ and matching them to
obtain the SR identity.

Based on the matrix elements of the hadrons $\mathrm{T}$, $B^{+}$ and $B^{-}$%
, the correlator $\Pi _{\mu \nu }(p,p^{\prime })$ is
\begin{eqnarray}
&&\Pi _{\mu \nu }^{\mathrm{Phys}}(p,p^{\prime })=\frac{\langle
0|J^{B^{+}}|B^{+}(p^{\prime })\rangle }{p^{\prime 2}-m_{B}^{2}}\frac{\langle
0|J^{B^{-}}|B^{-}(q)\rangle }{q^{2}-m_{B}^{2}}  \notag \\
&&\times \langle B^{+}(p^{\prime })B^{-}(q)|\mathrm{T}(p,\epsilon )\rangle
\frac{\langle \mathrm{T}(p,\epsilon )|I_{\alpha \beta }^{\dagger }|0\rangle
}{p^{2}-m^{2}}  \notag \\
&&+\cdots ,  \label{eq:PhenSide}
\end{eqnarray}%
where $m_{B}=(5279.42\pm 0.08)~\mathrm{MeV}$ is the mass of the mesons $%
B^{\pm }$ \cite{PDG:2024}.

The phenomenological side of the sum rule Eq.\ (\ref{eq:PhenSide}) is
approximated by a single pole term. In the case of the multiquark systems
this approach must be applied with some caution, because the physical side
receives contribution not only from the irreducible tetraquark-two meson
terms but also from two-meson reducible ones. In fact, the relevant
interpolating current couples not only to the tetraquark, but also to the
two-meson continuum lying below the mass of the multiquark system. In the
Feynman diagrams' language the irreducible contributions correspond to
connected diagrams whereas reducible terms are effects due to disconnected
diagrams.

This problem was emphasized while considering the sum rule for pentaquark
states \cite{Kondo:2004cr}, where the authors suggested to use the
two-hadron-irreducible correlation function, which is obtained by
subtracting reducible contributions from the standard correlation function.
Contributions of two-particle reducible terms were taken into account
explicitly in a phenomenological side of the sum rule for the pentaquark in
Refs.\ \cite{Lee:2004xk,Sarac:2005fn} and found around of $10\%$. In other
words, the reducible two-hadron contributions can be estimated in the
context of SR method itself by introducing corresponding couplings. In the
case of tetraquarks this approach was detailed and applied to compute strong
decays of various structures in Refs.\ \cite%
{Wang:2015nwa,Wang:2017lot,Wang:2019iaa,Wang:2021xao}.

For our purposes it is important that in all cases explored in Refs.\ \cite%
{Lee:2004xk,Sarac:2005fn,Wang:2017lot} and in other works, the two-hadron
continuum effects were found small and negligible. Moreover, explicit
computations of the width of the tetraquark $Z_{c}^{\pm}(3900)$ carried out
by taking into account only connected diagrams \cite{Dias:2013xfa} and both
the connected and disconnected ones \cite{Agaev:2016dev} gave very close
predictions. Therefore, in what follows, we derive the phenomenological side
of the sum rule by ignoring aforementioned subtleties.

We find $\Pi _{\mu \nu }^{\mathrm{Phys}}(p,p^{\prime })$ by means of matrix
elements
\begin{equation}
\langle 0|J^{B^{\pm }}|B^{\pm }\rangle =\frac{f_{B}m_{B}^{2}}{m_{b}},
\end{equation}%
and
\begin{equation}
\langle B^{+}(p^{\prime })B^{-}(q)|\mathrm{T}(p,\epsilon )\rangle
=g_{1}(q^{2})\epsilon _{\alpha \beta }(p)p^{\prime \alpha }p^{\prime \beta },
\end{equation}%
where $f_{B}=(206\pm 7)~\mathrm{MeV}$ is the decay constant of $B^{\pm }$.
After some manipulations, one gets
\begin{eqnarray}
&&\Pi _{\mu \nu }^{\mathrm{Phys}}(p,p^{\prime })=\frac{g_{1}(q^{2})\Lambda
f_{B}^{2}m_{B}^{4}}{m_{b}^{2}\left( p^{2}-m^{2}\right) \left( p^{\prime
2}-m_{B}^{2}\right) \left( q^{2}-m_{B}^{2}\right) }  \notag \\
&&\times \left[ \frac{m^{4}-2m^{2}(m_{B}^{2}+q^{2})+(m_{B}^{2}-q^{2})^{2}}{%
12m^{2}}g_{\mu \nu }\right.  \notag \\
&&\left. +p_{\mu }^{\prime }p_{\nu }^{\prime }+\text{other terms}\right] .
\end{eqnarray}%
It is clear that $\Pi _{\mu \nu }^{\mathrm{Phys}}(p,p^{\prime })$ contains
different Lorentz structures. For our analysis we use the terms $\sim g_{\mu
\nu }$ and denote by\ $\Pi _{1}^{\mathrm{Phys}}(p^{2},p^{\prime 2},q^{2})$
corresponding invariant amplitude.

For $\Pi _{\mu \nu }^{\mathrm{OPE}}(p,p^{\prime })$, we find%
\begin{eqnarray}
&&\Pi _{\mu \nu }^{\mathrm{OPE}}(p,p^{\prime })=-\frac{4}{3}\int
d^{4}xd^{4}ye^{ip^{\prime }y}e^{-ipx}\langle \overline{b}b\rangle  \notag \\
&&\times \left\{ \mathrm{Tr}\left[ \gamma _{5}{}S_{u}^{ij}(y)\gamma
_{5}S_{b}^{ja}(-x)\gamma _{\mu }\gamma _{\nu }S_{b}^{ai}(x-y)\right] \right.
\notag \\
&&\left. +\mathrm{Tr}\left[ \gamma _{5}{}S_{u}^{ij}(y)\gamma
_{5}S_{b}^{ja}(-x)\gamma _{\nu }\gamma _{\mu }S_{b}^{ai}(x-y)\right]
\right\} ,  \label{eq:QCDside1}
\end{eqnarray}%
where $S_{u}(x)$ is the $u$ quark's propagator \cite{Agaev:2020zad} and $%
\langle \overline{b}b\rangle $ is the vacuum matrix element of $\overline{b}%
b $. In following calculations, we utilize the relation
\begin{equation}
\langle \overline{b}b\rangle \approx -\frac{1}{12m_{b}}\langle \frac{\alpha
_{s}G^{2}}{\pi }\rangle
\end{equation}%
between the condensates $\langle \overline{b}b\rangle $ and $\langle \alpha
_{s}G^{2}/\pi \rangle $ obtained in Ref.\ \cite{Shifman:1978bx}. We label by
$\Pi _{1}^{\mathrm{OPE}}(p^{2},p^{\prime 2},q^{2})$ the amplitude that
corresponds in $\Pi _{\mu \nu }^{\mathrm{OPE}}(p,p^{\prime })$ to the
structure proportional to $g_{\mu \nu }$.

To extract the sum rule for $g_{1}(q^{2})$, we use the amplitudes $\Pi _{1}^{%
\mathrm{Phys}}(p^{2},p^{\prime 2},q^{2})$ and $\Pi _{1}^{\mathrm{OPE}%
}(p^{2},p^{\prime 2},q^{2})$ and get
\begin{eqnarray}
&&g_{1}(q^{2})=\frac{12m^{2}m_{b}^{2}(q^{2}-m_{B}^{2})}{\Lambda
f_{B}^{2}m_{B}^{4}[m^{4}-2m^{2}(m_{B}^{2}+q^{2})+(m_{B}^{2}-q^{2})^{2}]}
\notag \\
&&\times e^{m^{2}/M_{1}^{2}}e^{m_{B}^{2}/M_{2}^{2}}\Pi _{1}(\mathbf{M}^{2},%
\mathbf{s}_{0},q^{2}).
\end{eqnarray}%
Here, $\Pi _{1}(\mathbf{M}^{2},\mathbf{s}_{0},q^{2})$ is the amplitude $\Pi
_{1}^{\mathrm{OPE}}(p^{2},p^{\prime 2},q^{2})$ after the Borel
transformation and subtraction procedures: It can be expressed in terms of
the spectral density $\rho _{1}(s,s^{\prime },q^{2})$ calculated as an
imaginary part of relevant component of the correlation function $\Pi _{\mu
\nu }^{\mathrm{OPE}}(p,p^{\prime })$,
\begin{eqnarray}
&&\Pi _{1}(\mathbf{M}^{2},\mathbf{s}_{0},q^{2})=\int_{16m_{b}^{2}}^{s_{0}}ds%
\int_{m_{b}^{2}}^{s_{0}^{\prime }}ds^{\prime }\rho _{1}(s,s^{\prime },q^{2})
\notag \\
&&\times e^{-s/M_{1}^{2}}e^{-s^{\prime }/M_{2}^{2}},  \label{eq:SCoupl}
\end{eqnarray}%
where $\mathbf{M}^{2}=(M_{1}^{2},M_{2}^{2})$ and $\mathbf{s}%
_{0}=(s_{0},s_{0}^{\prime })$ are the Borel and continuum threshold
parameters, respectively. The couple of parameters $(M_{1}^{2},s_{0})$
corresponds to the channel of initial particle $\mathrm{T}$, whereas $%
(M_{2}^{2},s_{0}^{\prime })$ describes the channel of $B^{+}$ meson.

In computations for the $\mathrm{T}$ channel we employ parameters $%
(M_{1}^{2},s_{0})$ given by Eq.\ (\ref{eq:Wind1}). The parameters $%
(M_{2}^{2},s_{0}^{\prime })$ are varied within limits
\begin{eqnarray}
M_{2}^{2} &\in &[5.5,6.5]~\mathrm{GeV}^{2},\ s_{0}^{\prime }\in \lbrack
33.5,34.5]~\mathrm{GeV}^{2}.  \notag \\
&&
\end{eqnarray}

\begin{figure}[h]
\includegraphics[width=8.5cm]{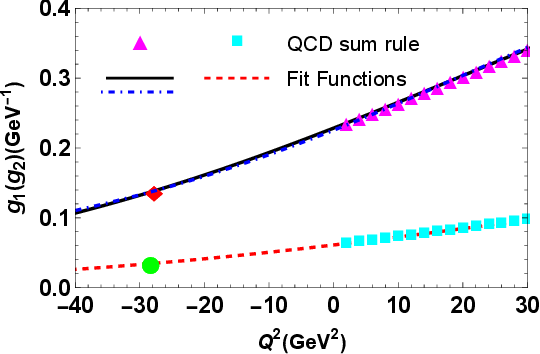}
\caption{ SR data and fit functions $\mathcal{Z}_1(Q^{2})$ (solid line), $%
\protect\widetilde{\mathcal{Z}}_1(Q^{2})$ (dot-dashed line) and $\mathcal{Z}%
_2(Q^{2})$ (dashed line) for the strong couplings $g_1$ and $g_2$. The
diamond and circle mark points $\mathcal{Z}_1(-m_{B}^{2})$ and $\mathcal{Z}%
_2(-m_{B^{\ast}}^{2})$, respectively.}
\label{fig:Fit}
\end{figure}

It is known that the sum rule method results in credible predictions for the
form factor $g_{1}(q^{2})$ in the deep Euclidean portion $q^{2}\ll 0$. In
the present analysis, we fix $-q^{2}=2-30~\mathrm{GeV}^{2}$ and show
obtained results in Fig.\ \ref{fig:Fit}. But the coupling $g_{1}$ has to be
extracted at the mass shell $q^{2}=m_{B}^{2}$. To solve this problem we
include into consideration the extrapolating function
\begin{equation}
\mathcal{Z}_{i}(Q^{2})=\mathcal{Z}_{i}^{0}\mathrm{\exp }\left[ z_{i}^{1}%
\frac{Q^{2}}{m^{2}}+z_{i}^{2}\left( \frac{Q^{2}}{m^{2}}\right) ^{2}\right] ,
\label{eq:FitF}
\end{equation}%
where $Q^{2}=-q^{2}$. Here, $\mathcal{Z}_{i}^{0}$, $z_{i}^{1}$ and $%
z_{i}^{2} $ are parameters which are chosen in such a way that $\mathcal{Z}%
_{i}(Q^{2})$ for $Q^{2}=2-30~\mathrm{GeV}^{2}$ coincides with outcomes of
the SR calculations. But it can be extrapolated also to domain of negative $%
Q^{2}$ which at $Q^{2}=-m_{B}^{2}$ gives the coupling $g_{1}$. This trick
allows us to estimate $g_{1}$
\begin{equation}
g_{1}\equiv \mathcal{Z}_{1}(-m_{B}^{2})=(1.38\pm 0.27)\times 10^{-1}\
\mathrm{GeV}^{-1}.  \label{eq:g1}
\end{equation}%
The fit function $\mathcal{Z}_{1}(Q^{2})$ used in these computations has the
parameters $\mathcal{Z}_{1}^{0}=0.229~\mathrm{GeV}^{-1}$, $z_{1}^{1}=5.423$,
and $z_{1}^{2}=-9.429$ and is plotted in Fig.\ \ref{fig:Fit} as well.

The result for $g_{1}$ has been found by employing Eq.\ (\ref{eq:FitF}). But
SR data can be extended into region the region $Q^{2}<0$ by means of
alternative extrapolating functions. To determine uncertainties of analysis
connected with such different choices, we consider the function
\begin{equation}
\widetilde{\mathcal{Z}}_{1}(Q^{2})=\frac{b_{0}}{\left( 1-Q^{2}/m^{2}\right)
^{3}[1-b_{1}(Q^{2}/m^{2})+b_{2}\left( Q^{2}/m^{2}\right) ^{2}]},
\label{eq:FitF2}
\end{equation}%
with $b_{0}$, $b_{1}$, and $b_{2}$ being fitted parameters. We estimate them
by comparing Eq. (\ref{eq:FitF2}) and SR data and find that the function
with parameters $b_{0}=0.226\ \mathrm{GeV}^{-1}$, $b_{1}=2.609$, and $%
b_{2}=11.923$ leads to very nice agreement with SR data. The function $%
\widetilde{\mathcal{Z}}_{1}(Q^{2})$ is plotted in Fig.\ \ref{fig:Fit} as
dot-dashed line and almost coincides with $\mathcal{Z}_{1}(Q^{2})$. Having
used $\widetilde{\mathcal{Z}}_{1}(Q^{2})$ we obtain $\widetilde{g}%
_{1}=0.1382~\mathrm{GeV}^{-1}$, which deviates from Eq.\ (\ref{eq:g1}) by an
amount of $0.0002$. \ This difference is two order of magnitude smaller than
ambiguities $\pm 0.027$ of the strong coupling due to the sum rule method.
Therefore, we neglect uncertainties appearing owing to different
extrapolating functions and apply Eq.\ (\ref{eq:FitF}) in  the present
article.

The width of the channel $\mathrm{T}\rightarrow B^{+}B^{-}$ is found by
means of the formula
\begin{equation}
\Gamma \left[ \mathrm{T}\rightarrow B^{+}B^{-}\right] =g_{1}^{2}\frac{%
\lambda _{1}}{960\pi m^{2}}(m^{2}-4m_{B}^{2})^{2},
\end{equation}%
where $\lambda _{1}=\lambda (m^{2},m_{B}^{2},m_{B}^{2})$ and
\begin{equation}
\lambda (a,b,c)=\frac{\sqrt{%
a^{4}+b^{4}+c^{4}-2(a^{2}b^{2}+a^{2}c^{2}+b^{2}c^{2})}}{2a}.
\end{equation}

The partial width of the channel $\mathrm{T}\rightarrow B^{+}B^{-}$ is equal
to
\begin{equation}
\Gamma \left[ \mathrm{T}\rightarrow B^{+}B^{-}\right] =(7.6\pm 2.2)~\mathrm{%
MeV}.
\end{equation}%
The width of the decay $\mathrm{T}\rightarrow B^{0}\overline{B}^{0}$ is
approximately equal to $\Gamma \left[ \mathrm{T}\rightarrow B^{+}B^{-}\right]
$.

\subsection{Decays $\mathrm{T}\rightarrow B^{\ast +}B^{\ast -}$,$\ B^{\ast 0}%
\overline{B}^{\ast 0}$}


In this part, we consider the decay mode $\mathrm{T}\rightarrow B^{\ast
+}B^{\ast -}$ in a detailed form. To extract the strong coupling $g_{2}$ at
the tetraquark-meson-meson vertex $\mathrm{T}B^{\ast +}B^{\ast -}$, we study
the correlation function%
\begin{eqnarray}
\Pi _{\mu \nu \alpha \beta }(p,p^{\prime }) &=&i^{2}\int
d^{4}xd^{4}ye^{ip^{\prime }y}e^{-ipx}\langle 0|\mathcal{T}\{J_{\mu
}^{B^{\ast +}}(y)  \notag \\
&&\times J_{\nu }^{B^{\ast -}}(0)I_{\alpha \beta }^{\dagger }(x)\}|0\rangle ,
\label{eq:CF1A}
\end{eqnarray}%
where $J_{\mu }^{B^{\ast +}}(x)$ and $J_{\nu }^{B^{\ast -}}(x)$ are the
interpolating currents of the mesons $B^{\ast +}$ and $B^{\ast -}$
\begin{equation}
J_{\mu }^{B^{\ast +}}(x)=\overline{b}_{i}(x)\gamma _{\mu }u_{i}(x),\ J_{\nu
}^{B^{\ast -}}(x)=\overline{u}_{j}(x)\gamma _{\nu }b_{j}(x).  \label{eq:CRB}
\end{equation}

In terms of the matrix elements of the states $\mathrm{T}$ , $B^{\ast +}$,
and $B^{\ast -}$ the correlator $\Pi _{\mu \nu \alpha \beta }(p,p^{\prime })$
is
\begin{eqnarray}
&&\Pi _{\mu \nu \alpha \beta }^{\mathrm{Phys}}(p,p^{\prime })=\frac{\langle
0|J_{\mu }^{B^{\ast +}}|B^{\ast +}(p^{\prime },\varepsilon _{1})\rangle }{%
p^{\prime 2}-m_{B^{\ast }}^{2}}\frac{\langle 0|J_{\nu }^{B^{\ast -}}|B^{\ast
-}(q,\varepsilon _{2})\rangle }{q^{2}-m_{B^{\ast }}^{2}}  \notag
\label{eq:CF2} \\
&&\times \langle B^{\ast +}(p^{\prime },\varepsilon _{1})B^{\ast
-}(q,\varepsilon _{2})|\mathrm{T}(p,\epsilon )\rangle \frac{\langle \mathrm{T%
}(p,\epsilon )|I_{\alpha \beta }^{\dagger }|0\rangle }{p^{2}-m^{2}}  \notag
\\
&&+\cdots ,
\end{eqnarray}%
where $m_{B^{\ast }}=(5324.75\pm 0.20)~\mathrm{MeV}$ is the mass of the
mesons $B^{\ast \pm }$, and $\varepsilon _{1\mu }$ and $\varepsilon _{2\nu }$
are their polarization vectors, respectively.

The function $\Pi _{\mu \nu \alpha \beta }^{\mathrm{Phys}}(p,p^{\prime })$
is obtained using the matrix elements
\begin{eqnarray}
\langle 0|J_{\mu }^{B^{\ast +}}|B^{\ast +}(p^{\prime },\varepsilon
_{1})\rangle &=&f_{B^{\ast }}m_{B^{\ast }}\varepsilon _{1\mu }(p^{\prime }),
\notag \\
\langle 0|J_{\nu }^{B^{\ast -}}|B^{\ast -}(q,\varepsilon _{2})\rangle
&=&f_{B^{\ast }}m_{B^{\ast }}\varepsilon _{2\nu }(q),  \label{eq:ME2B}
\end{eqnarray}%
with $f_{B^{\ast }}=(210\pm 6)~\mathrm{MeV}$ being the decay constant of $%
B^{\ast \pm }$ mesons \cite{Narison:2015nxh}. The vertex $\langle B^{\ast
+}(p^{\prime },\varepsilon _{1})B^{\ast -}(q,\varepsilon _{2})|\mathrm{T}%
(p,\epsilon )\rangle $ is given by the following expression
\begin{eqnarray}
&&\langle B^{\ast +}(p^{\prime },\varepsilon _{1})B^{\ast -}(q,\varepsilon
_{2})|\mathrm{T}(p,\epsilon )\rangle =g_{2}(q^{2})\epsilon _{\tau \rho }%
\left[ (\varepsilon _{1}^{\ast }\cdot q)\right.  \notag \\
&&\times \varepsilon _{2}^{\tau \ast }p^{\prime \rho }+(\varepsilon
_{2}^{\ast }\cdot p^{\prime })\varepsilon _{1}^{\ast \tau }q^{\rho
}-(p^{\prime }\cdot q)\varepsilon _{1}^{\tau \ast }\varepsilon _{2}^{\rho
\ast }  \notag \\
&&\left. -(\varepsilon _{1}^{\ast }\cdot \varepsilon _{2}^{\ast })p^{\prime
\tau }q^{\rho }\right] .  \label{eq:TVV}
\end{eqnarray}%
Then, for $\Pi _{\mu \nu \alpha \beta }^{\mathrm{Phys}}(p,p^{\prime })$ we
find
\begin{eqnarray}
&&\Pi _{\mu \nu \alpha \beta }^{\mathrm{Phys}}(p,p^{\prime })=g_{2}(q^{2})%
\frac{\Lambda f_{B^{\ast }}^{2}m_{B^{\ast }}^{2}}{\left( p^{2}-m^{2}\right)
(p^{\prime 2}-m_{B^{\ast }}^{2})}  \notag \\
&&\times \frac{1}{(q^{2}-m_{B^{\ast }}^{2})}\left[ p_{\beta }^{\prime
}p_{\alpha }^{\prime }g_{\mu \nu }+\frac{1}{2}p_{\mu }p_{\alpha }^{\prime
}g_{\beta \nu }\right.  \notag \\
&&\left. +\frac{1}{2m^{2}}p_{\beta }p_{\nu }p_{\mu }^{\prime }p_{\alpha
}^{\prime }+\text{ other terms}\right] +\cdots .
\end{eqnarray}

The sum rule for the strong form factor $g_{2}(q^{2})$ is obtained by
employing the amplitude $\Pi _{2}^{\mathrm{Phys}}(p^{2},p^{\prime 2},q^{2})$
that corresponds in $\Pi _{\mu \nu \alpha \beta }^{\mathrm{Phys}%
}(p,p^{\prime })$ to the term $\sim p_{\beta }p_{\nu }p_{\mu }^{\prime
}p_{\alpha }^{\prime }$.

The correlator $\Pi _{\mu \nu \alpha \beta }(p,p^{\prime })$ computed by
applying the quark propagators equals to%
\begin{eqnarray}
&&\Pi _{\mu \nu \alpha \beta }^{\mathrm{OPE}}(p,p^{\prime })=-\frac{4}{3}%
\int d^{4}xd^{4}ye^{ip^{\prime }y}e^{-ipx}\langle \overline{b}b\rangle
\notag \\
&&\times \left\{ \mathrm{Tr}\left[ {}\gamma _{\mu }S_{u}^{ij}(y)\gamma _{\nu
}S_{b}^{ja}(-x)\gamma _{\alpha }\gamma _{\beta }S_{b}^{ai}(x-y)\right]
\right.  \notag \\
&&\left. +\mathrm{Tr}\left[ \gamma _{\mu }S_{u}^{ij}(y)\gamma _{\nu
}S_{b}^{ja}(-x)\gamma _{\beta }\gamma _{\alpha }S_{b}^{ai}(x-y)\right]
\right\} .  \label{eq:QCDside2}
\end{eqnarray}

The sum rule for the strong form factor $g_{2}(q^{2})$ reads%
\begin{eqnarray}
&&g_{2}(q^{2})=\frac{2m^{2}(q^{2}-m_{B^{\ast }}^{2})}{\Lambda f_{B^{\ast
}}^{2}m_{B^{\ast }}^{2}}e^{m^{2}/M_{1}^{2}}e^{m_{B^{\ast
}}^{2}/M_{2}^{2}}\Pi _{2}(\mathbf{M}^{2},\mathbf{s}_{0},q^{2}).  \notag \\
&&
\end{eqnarray}%
In the $B^{\ast +}$ meson's channel we apply the parameters%
\begin{equation}
M_{2}^{2}\in \lbrack 5.5,6.5]~\mathrm{GeV}^{2},\ s_{0}^{\prime }\in \lbrack
34,35]~\mathrm{GeV}^{2}.  \label{eq:Wind2}
\end{equation}%
The coupling $g_{2}$ is evaluated using the SR data for $Q^{2}=2-30~\mathrm{%
GeV}^{2}$ (see, Fig. \ref{fig:Fit}) and extrapolating function with
parameters $\mathcal{Z}_{2}^{0}=0.016~\mathrm{GeV}^{-1}$, $z_{2}^{1}=6.136$,
and $z_{2}^{2}=-9.675$ which is shown in Fig.\ \ref{fig:Fit} as a dashed
line. The coupling $g_{2}$ is calculated at the mass shell $q^{2}=m_{B^{\ast
}}^{2}$ and equal to
\begin{equation}
g_{2}\equiv \mathcal{Z}_{2}(-m_{B^{\ast }}^{2})=(3.45\pm 0.66)\times
10^{-2}\ \mathrm{GeV}^{-1}.  \label{eq:G1}
\end{equation}%
The width of the decay $\mathrm{T}\rightarrow B^{\ast +}B^{\ast -}$ can be
found by means of the expression
\begin{eqnarray}
\Gamma \left[ \mathrm{T}\rightarrow B^{\ast +}B^{\ast -}\right] &=&g_{2}^{2}%
\frac{\lambda _{2}}{80\pi m^{2}}(m^{4}-3m^{2}m_{B^{\ast }}^{2}+6m_{B^{\ast
}}^{4}),  \notag \\
&&
\end{eqnarray}%
and is
\begin{equation}
\Gamma \left[ \mathrm{T}\rightarrow B^{\ast +}B^{\ast -}\right] =(9.8\pm
2.7)~\mathrm{MeV}.
\end{equation}

The difference between the decays $\mathrm{T}\rightarrow B^{\ast +}B^{\ast
-} $ and $\mathrm{T}\rightarrow B^{\ast 0}\overline{B}^{\ast 0}$ is encoded
in the masses of the final-state mesons. With nice accuracy we adopt $\Gamma %
\left[ \mathrm{T}\rightarrow B^{\ast +}B^{\ast -}\right] \approx \Gamma %
\left[ \mathrm{T}\rightarrow B^{\ast 0}\overline{B}^{\ast 0}\right] $.

\begin{figure}[h]
\includegraphics[width=8.5cm]{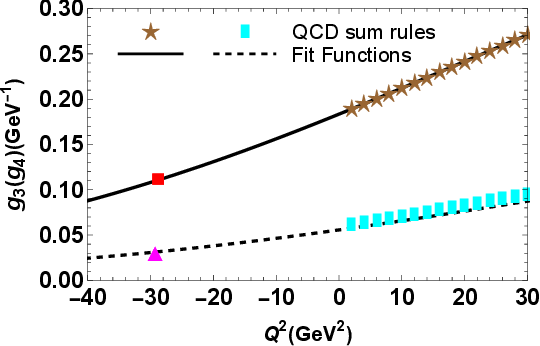}
\caption{ QCD data and extrapolating functions $\mathcal{Z}_3(Q^{2})$ (solid
curve) and $\mathcal{Z}_4(Q^{2})$ (dashed curve) for couplings $g_3$ and $%
g_4 $. The square and triangle denote positions $\mathcal{Z}_3(-m_{B_s}^{2})$
and $\mathcal{Z}_4(-m_{B_s^{\ast}}^{2})$.}
\label{fig:Fit1}
\end{figure}


\section{ Channels $\mathrm{T}\rightarrow B_{s}^{0}\overline{B}_{s}^{0}$ and
$\mathrm{T}\rightarrow B_{s}^{\ast 0}\overline{B}_{s}^{\ast 0}$}

\label{sec:Decays2}


Exploration of these processes does not differ considerably from studies of
the modes considered in the previous subsections: One has to take into
account some substitutions in the correlation functions and in parameters of
new final-state mesons. Indeed, the correlators of the channels $\mathrm{T}%
\rightarrow B_{s}^{0}\overline{B}_{s}^{0}$ and $\mathrm{T}\rightarrow
B_{s}^{\ast 0}\overline{B}_{s}^{\ast 0}$ can easily be obtained from Eqs.\ (%
\ref{eq:QCDside1}) and (\ref{eq:QCDside2}) after replacing $%
S_{u}^{ji}(y)\rightarrow S_{s}^{ji}(y)$.

Let us note that in computations we take into account terms $\sim
m_{s}=(93.5\pm 0.8)~\mathrm{MeV}$, but neglect ones proportional to $%
m_{s}^{2}$. Note that contributions $\sim m_{s}$ emerge in analytical
expressions owing to the propagator $S_{s}^{ji}(x)$ and matrix element
\begin{equation}
\langle 0|J^{B_{s}^{0}}|B_{s}^{0}\rangle =\frac{f_{B_{s}}m_{B_{s}}^{2}}{%
m_{b}+m_{s}}.
\end{equation}%
The spectroscopic parameters of the mesons $B_{s}^{0}$ and $B_{s}^{\ast 0}$
(as well as, $\overline{B}_{s}^{0}$ and $\overline{B}_{s}^{\ast 0}$ ) have
the following values
\begin{eqnarray}
m_{B_{s}} &=&(5366.93\pm 0.10)~\mathrm{MeV},\ f_{B_{s}}=(234\pm 5)~\mathrm{%
MeV},  \notag \\
m_{B_{s}^{\ast }} &=&(5415.4\pm 1.4)~\mathrm{MeV},\ f_{B_{s}^{\ast
}}=(221\pm 7)~\mathrm{MeV}.  \notag \\
&&
\end{eqnarray}

These processes are characterized by the strong couplings $g_{3}$ and $g_{4}$
at the vertices $\mathrm{T}B_{s}^{0}\overline{B}_{s}^{0}$ and $\mathrm{T}%
B_{s}^{\ast 0}\overline{B}_{s}^{\ast 0}$, respectively. The invariant
amplitudes $\Pi _{3(4)}(\mathbf{M}^{2},\mathbf{s}_{0},q^{2})$ utilized to
calculate the form factors $g_{3}(q^{2})$ and $g_{4}(q^{2})$ have forms
\begin{eqnarray}
&&\Pi _{3(4)}(\mathbf{M}^{2},\mathbf{s}_{0},q^{2})=%
\int_{16m_{b}^{2}}^{s_{0}}ds\int_{(m_{b}+m_{s})^{2}}^{s_{0}^{\prime
}}ds^{\prime }\rho _{3(4)}(s,s^{\prime },q^{2})  \notag \\
&&\times e^{-s/M_{1}^{2}}e^{-s^{\prime }/M_{2}^{2}}.
\end{eqnarray}%
The SR data for the form factor $g_{3}(q^{2})$ are calculated using the
working regions
\begin{equation}
M_{2}^{2}\in \lbrack 5.5,6.5]~\mathrm{GeV}^{2},\ s_{0}^{\prime }\in \lbrack
34,35]~\mathrm{GeV}^{2}.
\end{equation}%
The function $\mathcal{Z}_{3}(Q^{2})$ is evaluated by the fitted parameters
\begin{eqnarray}
\mathcal{Z}_{3}^{0} &=&0.185~\mathrm{GeV}^{-1},\ z_{3}^{1}=5.259,\
z_{3}^{2}=-9.138.  \notag \\
&&
\end{eqnarray}%
Then for the strong coupling $g_{3}$ we obtain the following prediction
\begin{equation}
g_{3}\equiv \mathcal{Z}_{3}(-m_{B_{s}}^{2})=(1.11\pm 0.20)\times 10^{-1}\
\mathrm{GeV}^{-1}.
\end{equation}%
The width of the process $\mathrm{T}\rightarrow B_{s}^{0}\overline{B}%
_{s}^{0} $ can be found via the expression%
\begin{equation}
\Gamma \left[ \mathrm{T}\rightarrow B_{s}^{0}\overline{B}_{s}^{0}\right]
=g_{3}^{2}\frac{\lambda _{3}}{960\pi m^{2}}(m^{2}-4m_{B_{s}}^{2})^{2},
\label{eq:PDw2}
\end{equation}%
where $\lambda _{3}=\lambda (m,m_{B_{s}},m_{B_{s}})$. As a result, we get
\begin{equation}
\Gamma \left[ \mathrm{T}\rightarrow B_{s}^{0}\overline{B}_{s}^{0}\right]
=(4.7\pm 1.3)~\mathrm{MeV}.  \label{eq:DW2}
\end{equation}

In the case of the decay $\mathrm{T}\rightarrow B_{s}^{\ast 0}\overline{B}%
_{s}^{\ast 0}$ parameters $(M_{2}^{2},\ s_{0}^{\prime })$ in the $%
B_{s}^{\ast 0}$ channel are changed within the borders
\begin{equation}
M_{2}^{2}\in \lbrack 6,7]~\mathrm{GeV}^{2},\ s_{0}^{\prime }\in \lbrack
35,36]~\mathrm{GeV}^{2}.
\end{equation}%
To estimate $g_{4}$, we have employed the extrapolation function $\mathcal{Z}%
_{4}(Q^{2})$ with parameters $\mathcal{Z}_{4}^{0}=0.056~\mathrm{GeV}^{-1}$, $%
z_{4}^{1}=5.983$, and $z_{4}^{2}=-9.403$.

The coupling $g_{4}$ amounts to
\begin{equation}
g_{4}\equiv \mathcal{Z}_{4}(-m_{B_{s}^{\ast }}^{2})=(3.12\pm 0.59)\times
10^{-2}\ \mathrm{GeV}^{-1}.
\end{equation}%
The width of this mode is equal to%
\begin{equation}
\Gamma \left[ \mathrm{T}\rightarrow B_{s}^{\ast 0}\overline{B}_{s}^{\ast 0}%
\right] =(7.9\pm 2.2)~\mathrm{MeV}.
\end{equation}%
The SR results for form factors $g_{3}(q^{2})$ and $g_{4}(q^{2})$ and
related extrapolating functions are drawn in Fig.\ \ref{fig:Fit1}.

Information gained in this and previous sections allow us to evaluate the
full decay width of the tensor tetraquark $\mathrm{T}$%
\begin{equation}
\Gamma \left[ \mathrm{T}\right] =(48\pm 6)~\mathrm{MeV}.
\end{equation}%
It turns out that $\mathrm{T}$ is not a very narrow state despite the fact
that is stable against fall-apart processes.


\section{Concluding notes}

\label{sec:Disc}

The fully-beauty diquark-antidiquark tensor state $\mathrm{T}=bb\overline{b}%
\overline{b}$ with the structure $C\gamma _{\mu }\otimes \gamma _{\nu
}C+C\gamma _{\nu }\otimes \gamma _{\mu }C$ considered in the present article
is an interesting object for investigations. First of all, one should
theoretically compute physical parameters of $\mathrm{T}$ by employing one
of known approaches of high energy physics. In our article, we have
calculated the mass and full decay width of this tetraquark in the framework
of QCD sum rule method. Our prediction $m=(18530\pm 86)~\mathrm{MeV}$ for
its mass proves that $\mathrm{T}$ does not decay to pairs of $\eta _{b}\eta
_{b}$ and $\Upsilon \Upsilon $ mesons. In other words, the tensor tetraquark
$\mathrm{T}$ cannot be seen as a peak in mass distributions of these
particles.

Parameters of the tensor tetraquark $bb\overline{b}\overline{b}$ were
calculated using different methods and presented in relevant publications.
Some of these results are collected in Table\ \ref{tab:Ten4bmasses}. As is
seen these predictions sometimes differ from each other considerbaly. The
QCD sum rule method used in this article and in Ref.\ \cite{Chen:2016jxd}
gives the identical predictions for the mass of the tetraquark $\mathrm{T}$.
Alternative methods, for example, the color-magnetic interaction approach
\cite{Wu:2016vtq}, or numerical analysis of the Schrodinger equation with
modified Cornell potential \cite{Likene:2026pga} generate considerably
larger results.

Different predictions for the mass lead to different decay patterns of the
tensor tetraquark $\mathrm{T}$. In some scenarios, the tensor state $\mathrm{%
T}$ can decay to $\eta _{b}\eta _{b}$ mesons, whereas in other ones it is
unstable against dissociation to both $\eta _{b}\eta _{b}$ and $\Upsilon
\Upsilon $ mesons. But in all options, it is important to find not only the
mass of a particle under discussion but also evaluate its width: Conclusions
drawn solely from the masses of these structures are not entirely convincing.

\begin{table}[tbp]
\begin{tabular}{|c|c|}
\hline\hline
Publications & Mass (in $\mathrm{MeV}$) \\ \hline
This work &$ 18530 \pm 86 $ \\
Ref.\ \cite{Berezhnoy:2011xn} & $18916$ \\
Ref.\ \cite{Wu:2016vtq} & $20243$ \\
Ref.\ \cite{Chen:2016jxd} & $18530 \pm 150$ \\
Ref.\ \cite{Wang:2017jtz} & $18850 \pm 90$ \\
Ref.\ \cite{Faustov:2022mvs} & $19331$ \\
Ref.\ \cite{Malekhosseini:2025hyx} & $17990 \pm 446$ \\
Ref.\ \cite{Likene:2026pga} & $19302$ \\ \hline\hline
\end{tabular}%
\caption{Predictions for the mass of the tensor tetraquark $bb\overline{b}%
\overline{b}$ made in some publications.}
\label{tab:Ten4bmasses}
\end{table}

In accordance to our results, the tetraquark $\mathrm{T}$ does not decay to $%
2\eta _{b}$ and $2\Upsilon $ mesons. We have evaluated the width of the
tetraquark $\mathrm{T}$ by exploring its six subleading decay modes. These
channels appear owing to annihilation of $b$-quarks in the tensor
tetraquark. Our prediction $\Gamma \left[ \mathrm{T}\right] =(48\pm 6)~%
\mathrm{MeV}$ for width of $\mathrm{T}$ is not large, therefore it may be
classified as a state with relatively moderate full width. Qualitatively
this means that the tensor structure $bb\overline{b}\overline{b}$ may be
observed in the mass distribution, for instance, of the meson pairs $%
B^{+}B^{-}$. But here one has to take into account and estimate possible
effects arising from dissociations of the scalar and tensor tetraquarks $bu%
\overline{b}\overline{u}$ and hadronic molecules $B^{+}B^{-}$. In other
words, there are various strong background noises to observe decays of $%
\mathrm{T}$.

The tetraquarks $bb\overline{b}\overline{b}$ till now are hypothetical
particles. But they can be observed in ongoing LHC experiments which have
certain potential to discover these structures \cite{Ali:2018xfq}. The Tera-$%
Z$ factory would have also such a potential \cite{Ali:2018ifm}. Therefore,
theoretical studies of tetraquarks $bb\overline{b}\overline{b}$ with
different spin-parities remain among important problems of particle physics.
This is especially true in light of the existing contradictory predictions
for their parameters. In the present article we have tried to illuminate one
of essential aspects of this problem.

\end{document}